\documentclass{WileyMSP-template}

\usepackage{graphicx}
\usepackage{amsmath}
\usepackage{amssymb}
\usepackage{bbold}
\usepackage[dvipsnames]{xcolor}
\usepackage{enumitem}
\usepackage[
    colorlinks, 
    citecolor=red,
    urlcolor=blue
]{hyperref}
\usepackage{url}
\usepackage{rotating}

\renewcommand{\d}{\delta}
\newcommand{\e}{\varepsilon}

\renewcommand{\th}{\theta}

\renewcommand{\k}{\kappa}

\newcommand{\m}{\mu}

\renewcommand{\t}{\tau}

\newcommand{\vc}[1]{\boldsymbol{\mathrm{#1}}}
\newcommand{\ts}[1]{\hat{\mathrm{#1}}}

\newcommand{\tsone}{\hat{\mathbb{1}}}
\newcommand{\diag}[1]{\operatorname{diag}\lrb{#1}}

\renewcommand{\div}{\nabla\cdot}

\renewcommand{\Im}{\mathrm{Im}}
\renewcommand{\Re}{\mathrm{Re}}

\newcommand{\lrp}[1]{\left(#1\right)}
\newcommand{\lrb}[1]{\left[#1\right]}

\newcommand{\lrv}[1]{\left|#1\right|}

\begin{document}
\pagestyle{fancy}

\title{Anomalous Reflection From Hyperbolic Media}
\maketitle


\author{Ilya Deriy*}
\author{Kseniia Lezhennikova}
\author{Stanislav Glybovksi}
\author{Ivan Iorsh}
\author{Oleh Yermakov}
\author{Mingzhao Song}
\author{Redha Abdeddaim}
\author{Stefan Enoch}
\author{Pavel Belov}
\author{Andrey Bogdanov*}


\dedication{This project was fulfilled during a collaboration dated 2019-2021}

\begin{affiliations}
Ilya Deriy$^{1\dagger}$, Kseniia Lezhennikova$^{2,3}$, Stanislav Glybovski$^{4}$, Ivan Iorsh$^{5}$,\\
 Oleh Yermakov$^{4}$, Mingzhao Song$^{1}$, Redha Abdeddaim$^{2}$, Stefan Enoch$^{2}$,\\ Pavel Belov$^{4}$, and Andrey Bogdanov$^{1,\dagger\dagger}$.
\\
\normalsize{$^{1}$Qingdao Innovation and Development Center, Harbin Engineering University,}
\\
\normalsize{Sansha Rd. 1777, Qingdao, Shandong, China, 266000}
\\
\normalsize{$^{2}$Aix Marseille Univ, CNRS, Centrale Marseille, Institut Fresnel}
\\
\normalsize{Institut Marseille Imaging, AMUTech, 13013 Marseille, France}
\\
\normalsize{$^{3}$Multiwave Technologies AG, 3 Chemin du Pré Fleuri 1228, Geneva, Switzerland}
\\
\normalsize{$^{4}$Independent researcher}
\\
\normalsize{$^{5}$Department of Physics, Engineering Physics and Astronomy, Queen’s University,}
\\
\normalsize{Kingston, Ontario K7L 3N6, Canada}
\\
\normalsize{$^\dagger$ideriy.physics@gmail.com}
\\
\normalsize{$^{\dagger\dagger}$a.bogdanov@hrbeu.edu.cn}
\end{affiliations}


\keywords{Hyperbolic Media; Anomalous Reflection; Electromagnetic Metamaterials}

\begin{abstract}
Despite the apparent simplicity, the problem of refraction of electromagnetic
waves at the planar interface between two media has an incredibly rich spectrum
of unusual phenomena. An example is the paradox that occurs when
an electromagnetic wave is incident on the interface between a hyperbolic
medium and an isotropic dielectric. At certain orientations of the optical axis
of the hyperbolic medium relative to the interface, the reflected and transmitted
waves are completely absent. In this paper, we formulate the aforementioned
paradox and present its resolution by introducing of infinitesimal losses
in the hyperbolic medium. We show that the reflected wave exists, but becomes
extremely decaying as the loss parameter tends to zero. As a consequence, all
the energy scattered into the reflected channel is absorbed near to interface.
We support our reasoning with analytical calculations, numerical simulations,
and an experiment with self-complementary metasurfaces in the microwave
range.
\end{abstract}


\section*{INTRODUCTION}
The refraction and reflection of electromagnetic waves at the planar interface between two media have been extensively studied for various systems, including nonlinear~\cite{bloembergen1962light}, anisotropic~\cite{lekner1991reflection}, chiral~\cite{lekner1996optical}, non-local media~\cite{luukkonen2009effects}, plasmonic and all-dielectric metasurfaces~\cite{yu2011light,paniagua2016generalized}.
The refraction of a plane electromagnetic wave at the interface is governed by the in-plane momentum conservation, also known as the phase-matching condition~\cite{jackson1999classical}. By knowing the dispersion of all propagating and evanescent waves at a certain frequency in both media, one can find amplitudes and directions of the refracted, reflected, and transmitted waves. The directions of the refracted, reflected, and transmitted waves can be easily found graphically using a technique of isofrequency contours (IFCs) --  the graphical solution of the dispersion equation at the certain frequency in the wave vector space (k-space) \cite{zengerle1987light, gralak2000anomalous}. The IFCs show all possible states of light in the medium at a certain frequency accessible for scattering.  The curvature and topology of IFCs crucially affect the structure of transmitted and reflected waves\cite{krishnamoorthy2012topological,orlov2014controlling}. The birefringence, negative refraction, and canalization can be easily understood in terms of IFCs~\cite{belov2005canalization,orlov2011engineered}.

In media with closed IFCs, electromagnetic waves can propagate in all directions, while in media with open IFCs, some propagation directions are forbidden which can result in unusual behavior of the reflected or transmitted waves~\cite{he2022anisotropy}. An illustrative example of a medium with an open isofrequency contour is a hyperbolic medium~\cite{poddubny2013hyperbolic}. A hyperbolic medium is a highly-anisotropic material that exhibits different signs of the real part of permittivity, permeability, or imaginary part of conductivity tensor components at the same frequency \cite{poddubny2013hyperbolic, ferrari2015hyperbolic, shekhar2014hyperbolic}.
Thus, the IFC of the hyperbolic media has the form of hyperboloids or hyperbolas in 3D and 2D cases, respectively~\cite{poddubny2013hyperbolic}.
The hyperbolic shape of the IFCs leads to the singular photonic density of states \cite{jacob2010engineering, poddubny2011spontaneous} which determines exotic properties of hyperbolic medium, such as negative refraction~\cite{argyropoulos2013negative,liu2013metasurfaces,high2015visible}, diffractionless~\cite{gomez2015hyperbolic,correas2017plasmon,yermakov2021surface} and high-directional~\cite{takayama2017midinfrared,gangaraj2019unidirectional,nemilentsau2019switchable} propagation, hybrid TE-TM polarization~\cite{yermakov2015hybrid,yermakov2018experimental}, and many more \cite{poddubny2013hyperbolic}.
These phenomena are widely used for a number of applications including the sub-diffraction imaging~\cite{fang2005sub,jacob2006optical}, in-plane hyperlensing~\cite{gomez2016flatland}, enhanced spontaneous emission~\cite{sreekanth2013directional,lu2014enhancing}, sensing~\cite{sreekanth2016extreme,sreekanth2016enhancing,shkondin2018high}, wavefront shaping and bending~\cite{yermakov2015hybrid,gomez2015hyperbolic,gomez2016flatland}, polarization and optical spin transformation~\cite{yermakov2016spin,mazor2020routing}, anomalous photonic spin Hall effect~\cite{kapitanova2014photonic,takayama2018photonic,kim2019observation}, and mimicking of plasma media~\cite{rustomji2018mimicking}.
Hyperbolic media can be implemented as layered metal-dielectric structures, nanorod arrays~\cite{poddubny2013hyperbolic,takayama2019optics,guo2020hyperbolic},
natural and microstructured two-dimensional materials~\cite{caldwell2014sub,nemilentsau2016anisotropic,dai2018manipulation,li2018infrared}, graphene-patterned structures~\cite{gomez2015hyperbolic,gomez2016flatland}, plasmonic gratings~\cite{high2015visible} and metasurfaces~\cite{yermakov2015hybrid,yermakov2018effective} in the visible and near-infrared spectra. Due to the open topology of the IFC, by properly aligning the hyperbolic medium's optical axis with respect to the interface with an isotropic dielectric, it is possible to obtain a regime in which the existence of the reflected wave is forbidden. The paradox of the absence of the reflected field was discussed in Ref.~\cite{guo2021abnormal}, and even accidentally observed in Ref.~\cite{dai2020edge} but it still has not found a resolution. 

In this work, we analyze and resolve the paradox of anomalous reflection from a hyperbolic media by showing that lossless approximation is incorrect for hyperbolic media without accounting for the usually neglected highly localized disappearing modes which we refer to as {\it ghost modes}. The ghost modes' manifestation becomes evident by introducing infinitely small losses. 
In the lossless case, the ghost modes have an infinitely high imaginary part of the propagation constant making them perfectly localized at the interface. Therefore, we predict that an interface of a hyperbolic medium cut at a certain orientation can serve as a perfect absorber for the waves incident at certain angles even if the material absorption in the bulk hyperbolic medium is negligible. We confirm our theoretical analysis numerically and experimentally in the GHz frequency range. Therefore, our insight into the physics of hyperbolic media allows us to resolve the paradox and predict the effect of broad angular perfect absorption in the vicinity of the hyperbolic media interface. 

\begin{figure}
    \centering
    \includegraphics[width=\linewidth]{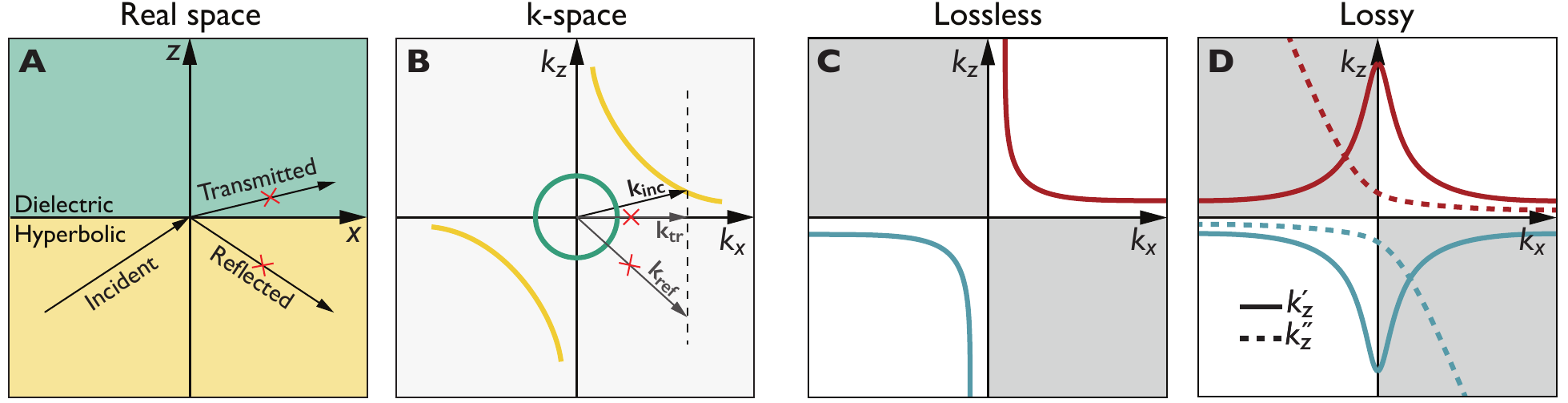}
    \caption{Schematic representation of the paradox that is the absence of reflected and transmitted waves in the case of refraction on the boundary between hyperbolic and isotropic media. \textbf A, \textbf B - schematic representation of the refraction of a plane electromagnetic wave on the boundary between hyperbolic and isotropic media in \textbf A - real space, \textbf B - k-space. When the hyperbolic medium is lossless, and when the modulus of the k-vector in the hyperbolic medium is larger than the modulus of the k-vector in the isotropic one, by rotation of the optical axis of the hyperbolic medium it is possible to create such a situation when there is no reflected and transmitted wave. \textbf C, \textbf D - schematic representation of the isofrequency contour for \textbf C - lossless and \textbf D - lossy hyperbolic medium.}
    \label{fig:main_idea}
\end{figure}

\section{Results}
\subsection{No-reflection paradox}
\label{sec:lossless}
\begin{figure}
    \centering
    \includegraphics[width=\linewidth]{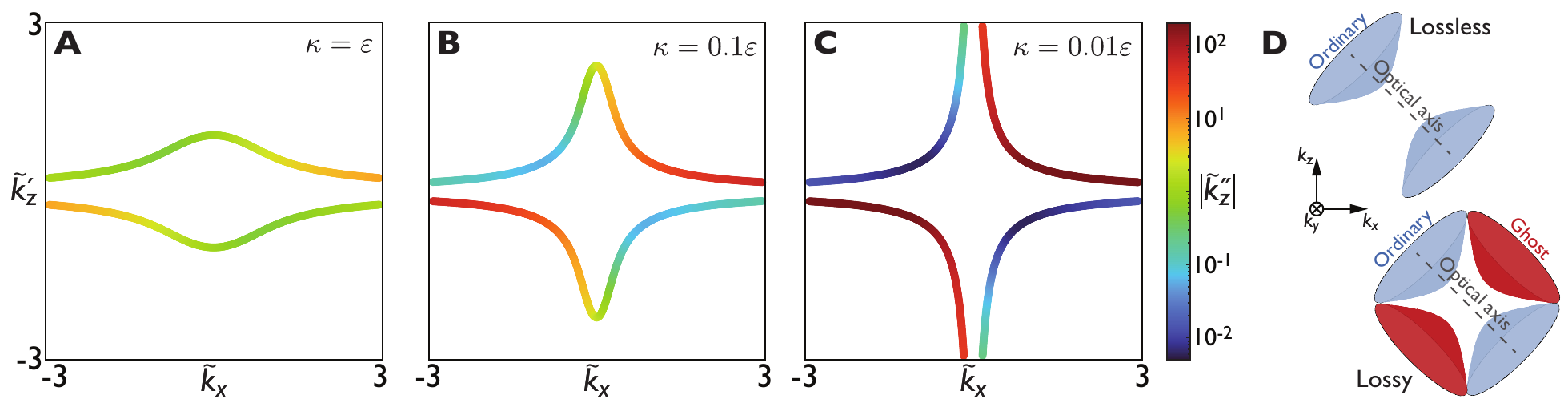}
    \caption{Isofrequency contours for lossy hyperbolic media with $\e = 1$ for different values of loss parameter $\k$, $\k = \e, \k = 0.1\e, \k = 0.01\e$. Colorbar encodes the modulus of the imaginary part of $\tilde k_z$. Panels \textbf A, \textbf B, and \textbf C show that as $\k\to0$, precisely in those quadrants in which the branches of the dispersion equation are absent for the lossless hyperbolic media, the imaginary part of $k_z$ is going to infinity, $\Im\lrb{k_z}\to\pm \infty$. Panel \textbf D shows schematical representation of the IFS for ordinary and ghost modes in lossless and lossy hyperbolic medium.}
    \label{fig:1_kvectors}
\end{figure}

Let us consider the Fresnel problem for the boundary between lossless hyperbolic medium and isotropic dielectric (Fig.~\ref{fig:main_idea}.\textbf A).
The permittivity tensor of the hyperbolic media in the principal axes is equal to $\hat \e = \mathrm{diag}\lrb{\e_o, \e_o, -\e_e}$. The optical axis of the hyperbolic medium is aligned in such a way that asymptotic lines of the dispersion are parallel to $k_x$ and $k_z$. All further results are applicable for the case of arbitrary $\e_o$ and $\e_e$ but, for the sake of simplicity, we put $\e_o=\e_e=\e$. Also, we draw the reader's attention to the fact that for brevity and clarity, where required, we will use the dimensionless components of the wave vector, which we mark with a tilde, and define as $\tilde k_{x,z} = k_{x,z}/k_0$. And, the real and imaginary parts of the quantities will be labeled with singe and double prime respectively (e.g. $\Re\lrb{k_z} \equiv k_z^\prime$, $\Im\lrb{k_z} \equiv k_{z}^{\prime\prime}$).  The relation between the wavevector components in such a hyperbolic media is equal to (see Sec.\ref{sec:analytical_derivaion})
\begin{equation}
    k_z = - \dfrac{1}{2}\dfrac{\e}{k_x}k_0^2.
    \label{eq:disp_lossless}
\end{equation}
Schematically, such an isofrequency contour is presented in  Figs.~\ref{fig:main_idea}.\textbf C,~\ref{fig:1_kvectors}.\textbf D.
In k-space, there is a branch of the isofrequency contour only in two quadrants.
Using the conservation of in-plane momentum it is possible to show that (for $k_x>0$) there is no IFC branch for the reflected wave.
In addition, if the wavevector in hyperbolic media is larger than the wavevector in an isotropic dielectric,  there will be no propagating transmitted wave as well.
As a consequence, both the reflected and transmitted waves will be absent in such a system (see Fig.~\ref{fig:main_idea}.\textbf A.).
Thus, there will be a constant energy inflow due to the presence of the incident wave and no channels for its consumption (neither for dissipation nor for scattering).
We call the aforementioned situation as \textit{no-reflection paradox}.

\subsection{Effect of losses}
\label{sec:resolution}
The resolution of the paradox lies in the introduction of losses to the system.
With the presence of isotropic losses $\kappa$, the permittivity tensor of the hyperbolic medium reads as
\begin{equation}
    \hat \e = \begin{bmatrix}
        i \k & 0 & \e
        \\
        0 &  i \k + \e & 0
        \\
        \e & 0 & i\k
    \end{bmatrix},
\end{equation}
where $\e,\k\in\mathbb R$.
It is important to stress that we have chosen $e^{- i \omega t}$ time dependence, and thus $\kappa>0$ corresponds to lossy media, while $\k<0$ corresponds to media with gain.
For such a permittivity tensor, the relation between wavevector components is given by
\begin{equation}
    k_{z_\pm} = \dfrac{i}{\k}\lrb{k_x \e \pm\sqrt{\lrp{k_x^2 - i \k k_0^2}\lrp{\k^2 + \e^2}}}.
    \label{eq:2_disp_complex}
\end{equation}
Figure \ref{fig:1_kvectors} shows the real part of $k_{z_\pm}$ from Eq.~\eqref{eq:2_disp_complex} as a function of $k_x$ for a hyperbolic medium with $\varepsilon = 1$ for different values of $\kappa$.
One can clearly see that in the lossy case, even for infinitesimal losses, there exists a branch of the dispersion curve in each quadrant of the k-space.
It is important to stress that for the same $k_x$ upper and lower branches of isofrequency contour, i.e. those that are usually associated with $k_{z_\mathrm{inc}}$ and $k_{z_\mathrm{ref}}$ are no longer related as $k_{z_\mathrm{inc}} = - k_{z_\mathrm{ref}}$ as it is usually taken.
To be precise, the latter identity holds, but only for the real part of $k_z$. The imaginary part, on the other hand, is decreasing for one branch and increasing for another one as $\kappa \to 0$ (see Fig. \ref{fig:1_kvectors}.\textbf A-c).
A much clearer picture can be obtained with the help of power expansion of Eq.~\eqref{eq:2_disp_complex} around $\k\approx 0$ that is
\begin{equation}
    k_{z_{\pm}} = \pm\dfrac{1}{2}\lrv{\dfrac{\e}{k_x}}k_0^2 + \dfrac{i}{\k}\lrp{k_x \e \pm \lrv{k_x \e}} \pm i\dfrac{\lrv{k_x \e}}{2}\lrp{\dfrac{k_0^4}{4 k_x^2}+ \dfrac{1}{\e^2}} \k + O(\k^2).
\end{equation}
The latter relation can be decomposed into two parts,
\begin{equation}
	k_z^\mathrm{o} = \underbrace{- \dfrac{1}{2} \dfrac{\e}{k_x} k_0^2}_{\substack{\text{lossless} \\ \text{dispersion}}} - \underbrace{i 
	\k \dfrac{k_x \e}{2} \lrp{\dfrac{k_0^4}{4 k_x^4} + \dfrac{1}{\e^2}} }_{\substack{\text{regular} \\ \text{ losses}}}, \quad k_z^\mathrm{g} =  \dfrac{1}{2} \dfrac{\e}{k_x} k_0^2+ \underbrace{i 
	\dfrac{2k_x \e}{\k}  }_{\substack{\text{singular} \\ \text{ losses}}},
	\label{eq:disp_lossy}
\end{equation}
where indices $"\mathrm{o}"$ and $"\mathrm{g}"$ denote words "ordinary" and "ghost".
The resolution of the no-reflection paradox lies in Eq.~\eqref{eq:disp_lossy}.
The real part of the dispersion for ordinary waves is identical to the dispersion in lossless hyperbolic media Eq.~\eqref{eq:disp_lossless}.
The real part of the dispersion for ghost waves shares the same modulus with propagating waves but has the opposite sign.
Therefore, dispersion for decaying waves is present in the quadrants which are empty in the lossless case (see Fig.~\ref{fig:main_idea}.\textbf D, Fig.~\ref{fig:1_kvectors}).
The imaginary part of the dispersion for ordinary waves is proportional to the loss parameter $\kappa$, which is common for any isotropic lossy medium.
However, the imaginary part of the dispersion for ghost waves is inversely proportional to the loss parameter, which follows the counter-intuitive conclusion that the smaller the loss, the higher the decaying rate of the ghost waves.
Thus, one can formulate the resolution of the paradox: \textit{the reflected wave is present in the system, but is highly decaying.}
It is important to stress that because as $\k \to 0$ imaginary part of $k_z^\mathrm{g}$ diverges, it is not always correct to consider hyperbolic media as lossy since the transition from lossless to lossy media is non-regular.

\begin{figure}
    \centering
    \includegraphics[width=\linewidth]{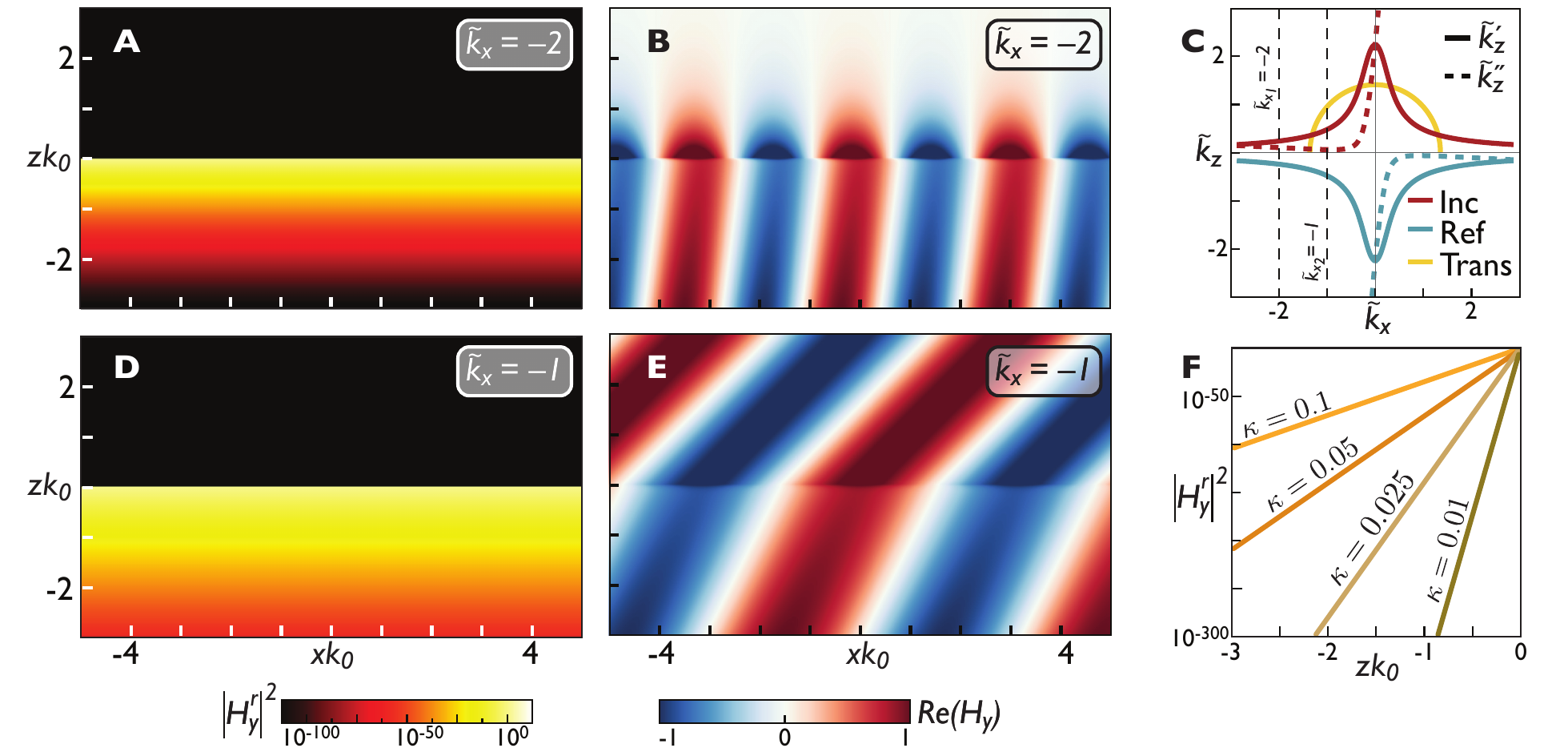}
    \caption{Refraction of a plane electromagnetic wave on the interface between lossy hyperbolic media with $\e = 1$ and isotropic dielectric with $\e_0 = 2$ for two different values of $\tilde k_x = k_x/k_0$ and several values of $\k$. \textbf A, \textbf D - modulus squared of reflected magnetic field. \textbf B, \textbf E - the real part of the total magnetic field. Panels \textbf A, \textbf B correspond to $\tilde k_x = -2$, panels \textbf D, \textbf E to $\tilde k_x = -1$. Panel \textbf C shows the k-space. All panels except \textbf F are plotted for $\k = 0.1$. Panel \textbf F shows the modulus squared of the reflected magnetic field for $\tilde k_x = -2$ and different values of $\k$.}
    \label{fig:fields}
\end{figure}
Figure \ref{fig:fields} shows refraction of a plane electromagnetic wave on the interface between lossy hyperbolic medium with $\e = 1$ and isotropic dielectric with $\e_0 = 2$ for two different values of $\tilde k_x$ and several values of $\k$. 
We present in this figure the only component of the magnetic field for TM wave ($H_y$).
One can clearly see that the reflected field is highly decaying (see Figs.~\ref{fig:fields}.\textbf A, \ref{fig:fields}.\textbf C, \ref{fig:fields}.\textbf F).
Additional reasoning can be presented to prove the highly decaying nature of the reflected field.
When amplitudes of the incident and reflected waves are comparable, one can observe the appearance of nodes and antinodes as a result of their interference.
From the absence of the interference pattern in the hyperbolic medium region, one can conclude that the amplitude of the reflected wave is almost negligible (see Figs.~\ref{fig:fields}.\textbf B, \ref{fig:fields}.\textbf D).

\begin{figure}
    \centering
    \includegraphics[width=\linewidth]{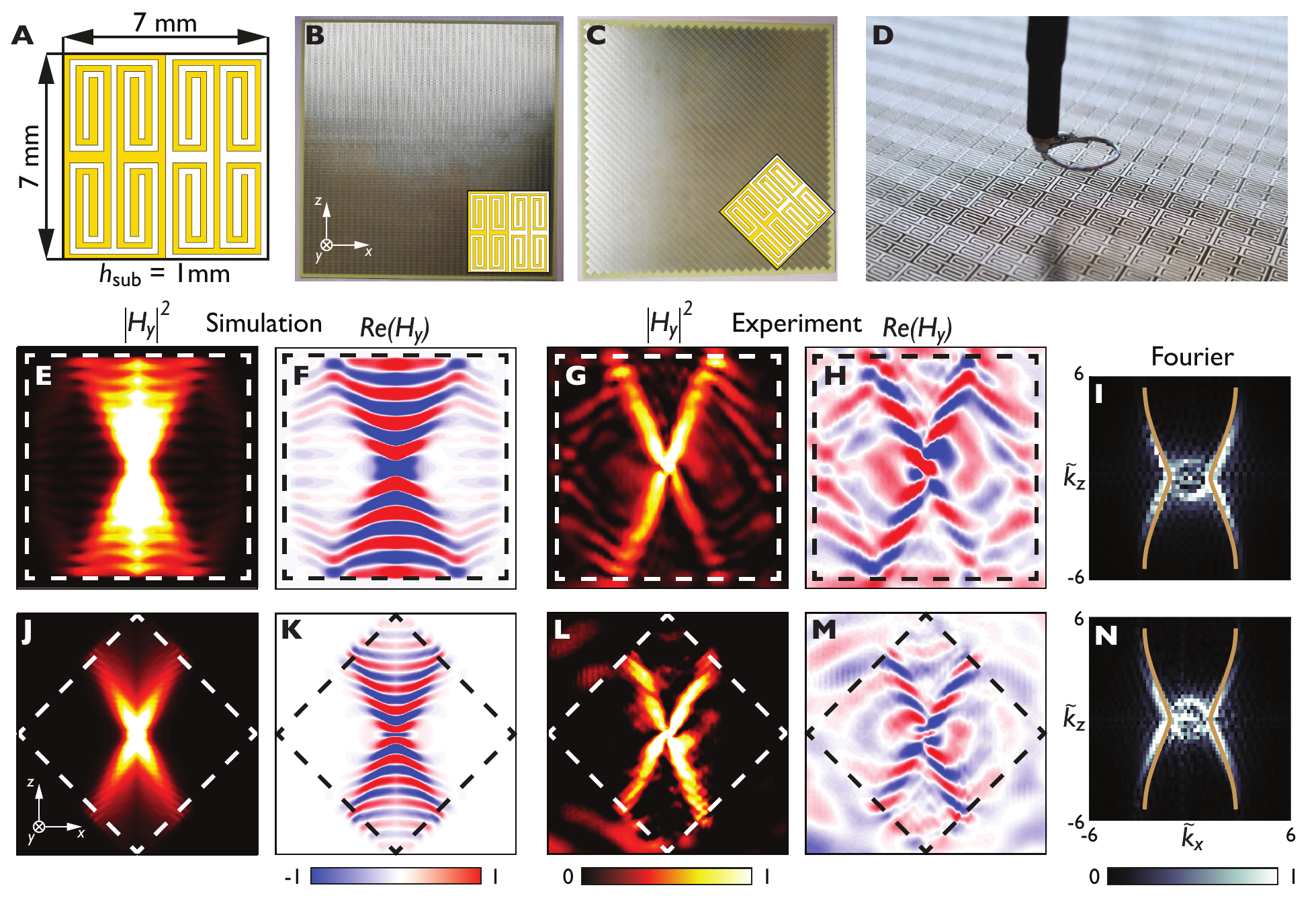}
    \caption{Numerical simulation and experimental results. \textbf A - unit cell of the metasurface. \textbf B, \textbf C - square and rhombic samples respectively. The insets shows the orientation of the unit cell. \textbf D - photo of the excitation loop over a metasurface. \textbf E-\textbf I, and \textbf J-\textbf N - numerical and experimental results for rhombic and square samples respectively. \textbf E, \textbf G, \textbf J, \textbf L, and \textbf F, \textbf H, \textbf K, \textbf M - modulus and real parts of $H_y$, respectively. Dashed rectangules represent the edge of the sample. \textbf I, \textbf N - two-dimensional Fourier transform of the experimentally measured normal component of the magnetic field, for the square and rhombic samples respectively, showing the hyperbolic dispersion of the metasurface. Brown lines correspond to the numerically calculated IFCs using the Eigenfrequency Solver of CST Microwave Studio.
    }
    \label{fig:experimental_fields}
\end{figure}
\subsection{Experimental verification}
For the numerical simulations and experiment we used a self-complementary hyperbolic metasurface (a 2D analog of a hyperbolic medium with a period of 7 mm) with unit cells designed exactly as in the work \cite{yermakov2021surface}. An FR-4 substrate with the relative permittivity $\e= 4.3$, loss tangent $\tan \d= 0.025$, and thickness $h=1\:\text{mm}$ was used to obtain the hyperbolic regime at 4 GHz (see Figs.~\ref{fig:experimental_fields}.\textbf I,~\ref{fig:experimental_fields}.\textbf N). The two samples have been designed, having dimensions of 280 mm $\times$ 280 mm and containing $40\times 40$ unit cells, i.e. the square sample (Fig. ~\ref{fig:experimental_fields} b) and the rhombic one (Fig. ~\ref{fig:experimental_fields} c) (with unit cells rotated by $45^{\circ}$ relative to ones of the square sample).
The excitation loop (Fig. ~\ref{fig:experimental_fields} d) of the diameter $d=7 \text{ mm}$, acting as the vertical magnetic dipole was placed in the middle of the sample at a distance of $4\:\text{mm}$ from the surface of the sample. A similar loop was used to measure the spatial distribution of the normal component of the magnetic field $H_y$. After measuring the spatial distribution we applied the two-dimensional Fourier transform to extract the IFCs as it was done in Refs.~\cite{yermakov2018experimental, yermakov2021surface}. The same transform was done for the numerically calculated field maps obtained using CST Microwave Studio. For more details on the numerical simulation and experiment see Sec.~\ref{sec:experimental_verification}. Results of the numerical simulation and the experiment, for both square and rhombic structures, are presented in Fig.~\ref{fig:experimental_fields}.
To prove the decaying nature of the reflected wave, one can resort to the same reasoning as in Sec.~\ref{sec:resolution}.
Figures~\ref{fig:experimental_fields}.\textbf E, and~\ref{fig:experimental_fields}.\textbf G show the absolute value of numerically and experimentally measured magnetic field for the square structure.
One can see the presence of pronounced fringes due to the interference of incident and reflected waves.
However, as discussed in Sec.~\ref{sec:resolution}, when asymptotes of the dispersion contour are perpendicular to the edge of the hyperbolic metasurface, the reflected wave becomes highly decaying.
As a result, the amplitude of the reflected wave is negligible in comparison to the amplitude of the incident wave, therefore, the interference pattern is absent for the rhombic structure (see Figs.~\ref{fig:experimental_fields}.\textbf J,~\ref{fig:experimental_fields}.\textbf L).
However, it is very important to clarify that since the dipole emits waves on all wave vectors, in numerical modeling and experiment they are present as waves completely reflected from the boundaries of the sample (the case corresponding to Figs.~\ref{fig:fields}.\textbf A,~\ref{fig:fields}.\textbf B), and waves flowing into free space (the case corresponding to Figs.~\ref{fig:fields}.\textbf D,~e\ref{fig:fields}.\textbf E).

\section{Discussion}
The results obtained show that hyperbolic medium with properly modeled losses leads to the presence of the isofrequency contour brances in each quadrant of the k-space.
However, as $\k\to 0$, precisely in those quadrants in which the branches of the dispersion curve are absent for the lossless hyperbolic media, the imaginary part of $k_z$ is diverging (Fig.~\ref{fig:1_kvectors}).
Thus it can be concluded that the transition from lossy to lossless hyperbolic media is non-regular, therefore it is important to always consider hyperbolic media as lossy.
It is important to stress that the normal component of the k-vector for the incident and reflected waves are no longer related as $k_z^i = - k_z^r$, as it is commonly considered.
To be precise, for small values of $\k$, $k_z^{i\prime} = - k_z^{r\prime}$, but $k_z^{i\prime\prime} \sim \k k_x$, while $k_z^{r\prime\prime} \sim k_x/\k$.
For small losses, while the incident wave is almost non-decaying, the reflected wave is decaying as $e^{k_x z/\k}$.

In the experiment, it was shown that the presence of the pronounced reflection from the edge of a hyperbolic metasurface strongly depends on the orientation of the edge relative to the axes of the hyperbolic dispersion. The case in which the propagation of the refracted waves is forbidden corresponds to negligible reflection. This was clearly observed as there was no interference between the incident and the reflected waves. As stated above this effect can be explained by a fast decay caused by moderate losses in the FR4 material and copper unit cells in the microwave range.

However, an important issue arises.
The theory described in this article can be applied to the considered metasurface only when considered as a homogenized two-dimensional medium.
Homogenization theory is usually valid when electromagnetic fields do not change much within the unit cell of the periodic structure.
However, as $\k\to0$, the reflected field becomes highly localized, due to which homogenization theory for the reflected field may fail.
In addition, it is important to stress, that while in the paper it is assumed that $\kappa >0$, which corresponds to losses, all equations are applicable for $\kappa<0$ too.
From this, one can conclude that the reflected field will experience extremely large field amplification in the case of hyperbolic media with gain even for small values of $\k$.

\section{Conclusions}
In this paper, we formulated a solution to the paradox of the absence of a reflected wave in hyperbolic metamaterials and metasurfaces for a certain orientation of the isofrequency contour relative to the interface between the hyperbolic and isotropic media. According to analytical calculations, the isofrequency contour for the reflected wave appears when losses are added to the hyperbolic medium. In this case, the reflected wave exists, but it has a strongly decaying nature, and the more decaying, the smaller the loss factor $\kappa$. Analytical calculations were confirmed by numerical simulation and experiment. It was shown in modeling and experiments that if the isofrequency contour of a hyperbolic medium is perpendicular to the interface, then the interference between the incident and reflected waves is almost negligible.
We believe that the results of our work will become the basis for the creation of perfect absorbers and perfect transmitters based on hyperbolic media.


\section{Materials and Methods}
\subsection{Analytical derivation}
\subsubsection{Dispersion relation for the lossless and lossy hyperbolic media}
\label{sec:analytical_derivaion}
Let us consider nonmagnetic uniaxial media described by the following permittivity tensor,
\begin{equation*}
    \hat\e = 
    \begin{bmatrix}
    \e_{xx} & 0 & \e_{xz}
    \\
    0 & \e_{yy} & 0
    \\
    \e_{zx} & 0 & \e_{zz}
    \end{bmatrix}.
\end{equation*}
The dispersion equation for electromagnetic waves in such media can be found by solving the following equation
\begin{equation}
    \det\left[k^2\tsone - \vc k\otimes \vc k-\hat\e k_0\right] = 0,
    \label{eq:disp_main}
\end{equation}
where $\vc k = [k_x, k_y, k_z]^T$ is the wavevector, and $k_0$ is the wavenumber in the free-space.
Under the assumption $k_y = 0$, Eq.~\eqref{eq:disp_main} factorizes and gives the dispersion of TE- and TM-polarized waves
\begin{equation}
   \underbrace{\left[k_x^2 + k_z^2 - \e_{yy} k_0^2\right]}_{\mathrm{TE}}\underbrace{\left[\e_{xx}k_x^2 + (\e_{xz} + \e_{zx})k_x k_z+ \e_{zz}k_z^2-\eta k_0^2\right]}_{\mathrm{TM}} = 0,
   \label{eq:1_dispeq}
\end{equation}
where $\eta = \operatorname{Det}\hat \e_\mathrm{2D}$,
\begin{equation}
    \hat \e_\mathrm{2D} = \begin{bmatrix}
        \e_{xx} & \e_{xz}
        \\
        \e_{zx} & \e_{zz}
    \end{bmatrix}.
\end{equation}
Thus, for TM-polarized wave, one can derive
\begin{equation}
    k_{z_\pm} = \dfrac{\pm \sqrt{4 \e_{zz} \eta k_0^2 + \lrp{\lrp{\e_{xz} + \e_{zx}}^2 - 4 \e_{xx}\e_{zz}} k_x^2} - \lrp{\e_{xz} + \e_{zx}} k_x }{2 \e_{zz}}.
    \label{eq:4_kzkx}
\end{equation}

In the case of lossless hyperbolic media with permittivity tensor $\hat \e = \diag{\e, \e, -\e}$, $\e \in \mathbb R$, rotation of the material by $\pi/4$ will result in the following transformation of $\hat \e$
\begin{equation}
    \hat\e = 
    \ts R\lrp{\pi/4}^\mathrm{T}
    \begin{bmatrix}
    \e & 0 & 0
    \\
    0 & \e & 0
    \\
    0 & 0 & -\e
    \end{bmatrix}
    \ts R\lrp{\pi/4} = 
    \begin{bmatrix}
    0 & 0 & \e
    \\
    0 & \e & 0
    \\
    \e & 0 & 0
    \end{bmatrix},
    \label{eq:1_rotated}
\end{equation}
where 
\begin{equation}
    \ts R\lrp{\theta} =
    \begin{bmatrix}
    \cos\th & 0 & \sin\th
    \\
    0 & 1 & 0
    \\
    -\sin\th & 0 & \cos\th
    \end{bmatrix}.
\end{equation}
Substitution of Eq.~\eqref{eq:1_rotated} into Eq.~\eqref{eq:4_kzkx} will give the following relation between $k_z$ and $k_x$ components of the wavevector of the TM-polarized wave
\begin{equation}
    k_z = -\dfrac{1}{2}\dfrac{\e}{k_x}k_0^2.
\end{equation}
One can see that isofrequency contours in this case are hyperbolas.
However, if losses are present in the system, the relation between $k_z$ and $k_x$ is changed drastically.
Let us introduce x arbitrary losses to the hyperbolic medium, $\pm\e \to \pm\e + i \k$, $\e,\k\in\mathbb R$.
Then,
\begin{equation}
    \hat\e = 
    \ts R\lrp{\pi/4}^\mathrm{T}
    \begin{bmatrix}
    \e+i\k & 0 & 0
    \\
    0 & \e+i\k & 0
    \\
    0 & 0 & -\e+i\k
    \end{bmatrix}
    \ts R\lrp{\pi/4} = 
    \begin{bmatrix}
    i\k & 0 & \e
    \\
    0 & \e + i \k & 0
    \\
    \e & 0 & i \k,
    \end{bmatrix}.
    \label{eq:1_rotated_loss}
\end{equation}
In this case the relation between $k_x$ and $k_z$ components is
\begin{equation}
    k_{z_\pm} = \dfrac{i}{\k}\left[k_x\e\pm\sqrt{(k_x^2-i\k k_0^2)(\k^2+\e^2)}\right].
\end{equation}

\subsubsection{Refraction coefficients for the interface between uniaxial and isotropic media.}
Let us consider the problem of refraction of a plane electromagnetic wave at the interface between uniaxial media with material parameters
\begin{equation}
    \hat\e = 
    \begin{bmatrix}
        \e_{xx} & 0 & \e_{xz}
        \\
        0 & \e_{yy} & 0
        \\
        \e_{zx} & 0 & \e_{zz}
    \end{bmatrix},
\end{equation}
$\m=1$, and isotropic dielectric with material parameters $\e_0$, $\m_0=1$.
We are interested in the refraction of the TM-polarized wave.
The coordinate system is chosen in such a way that the electric field of the incident wave lies in the $xz$ plane.
Boundary conditions for the interface between media "1" and media "2" (if there are no surface charges and currents) are
\begin{equation}
    \begin{aligned}
        &E_\t^{(1)} - E_\t^{(2)}=0, &H_\t^{(1)} - H_\t^{(2)}=0,
        \\
        &D_n^{(1)} - D_n^{(2)}=0, &B_n^{(1)} - B_n^{(2)}=0,
    \end{aligned}
\end{equation}
where $\t$ and $n$ denote transversal and normal, with respect to the interface, and components of fields.
In Cartesian coordinates, boundary conditions for the electric field will have the following look
\begin{equation}
    E_x^i+E_x^r = E_x^t,
    \label{eq:4_bc_el}
\end{equation}
and
\begin{equation}
    D_z^i + D_z^r = D_z^t.
    \label{eq:4_bc_disp}
\end{equation}
Under substitution of the material equation $\vc D= \hat \e \vc E$, Eq.~\eqref{eq:4_bc_disp} rewrites as
\begin{equation}
    \lrp{\e_{zx}E_x^i + \e_{zz}E_z^i} + \lrp{\e_{zx}E_x^r + \e_{zz}E_z^r} = \e_0 E_z^t.
    \label{eq:4_r_1}
\end{equation}
Using the Gauss law $\div \vc D = 0$, it is possible to show that
\begin{equation}
    E_z = - \dfrac{k_x \e_{xx} + k_z \e_{zx}}{k_x \e_{xz} + k_z \e_{zz}} E_x.
\end{equation}
Therefore, Eq.~\eqref{eq:4_r_1} can be simplified as follows
\begin{equation}
    \lrp{\e_{zx} - \e_{zz}\dfrac{k_x \e_{xx}+ k_z^i \e_{zx}}{k_x\e_{xz}+ k_z^i \e_{zz}}}E_x^i + \lrp{\e_{zx} - \e_{zz}\dfrac{k_x \e_{xx}+ k_z^r \e_{zx}}{k_x\e_{xz}+ k_z^r \e_{zz}}}E_x^r = - \e_0\dfrac{k_x}{k_z^t}E_x^t,
    \label{eq:4_r_2}
\end{equation}
where we used the fact that isotropy of space in x direction implies that momentum in that direction is conserved, i.e.
\begin{equation}
    k_x^i = k_x^r = k_x^t=k_x.
    \label{eq:4_cons_mom}
\end{equation}
Substitution of $E_x^t$ from Eq.~\eqref{eq:4_bc_el} into Eq.~\eqref{eq:4_r_2} gives following equation
\begin{equation}
    \lrp{\e_{zx} + \e_0 \dfrac{k_x}{k_z^t} - \e_{zz}\dfrac{k_x \e_{xx}+ k_z^i \e_{zx}}{k_x\e_{xz}+ k_z^i \e_{zz}} }E_x^i = \lrp{ \e_{zz}\dfrac{k_x \e_{xx}+ k_z^r \e_{zx}}{k_x\e_{xz}+ k_z^r \e_{zz}} - \e_{zx} - \e_0 \dfrac{k_x}{k_z^t}}E_x^r.
\end{equation}
Thus, the reflection coefficient for the tangential component of the electric field of the TM-polarized wave, after simplification, is
\begin{equation}
    r_x \equiv \dfrac{E_x^r}{E_x^i} = \dfrac{\lrp{\e_{xz}k_x + \e_{zz}k_z^r}\lrp{\eta k_z^t - \e_0\lrp{\e_{xz}k_x + \e_{zz}k_z^i}}}{\lrp{\e_{xz}k_x + \e_{zz}k_z^i}\lrp{\e_0\lrp{\e_{xz}k_x + \e_{zz}k_z^r} -\eta k_z^t} }.
\end{equation}
From boundary conditions it is clear that the transmittion coefficient is connected with the reflection coefficient as
\begin{equation}
    t_x = r_x + 1.
\end{equation}
It can be more convenient to solve the refraction problem via the only component of the magnetic field, thus we also derive the expression for reflection and transmission coefficients for the magnetic field.
From Maxwell's equations one can obtain
\begin{equation}
    \dfrac{k_z}{k_0}H_y = \e_{xx} E_x + \e_{xz} E_z,\quad - \dfrac{k_x}{k_0}H_y = \e_{zx}E_x + \e_{zz} E_z.
\end{equation}
from which it is possible to show that
\begin{equation}
    E_x = \dfrac{\e_{xz} k_x + \e_{zz} k_z}{\eta} H_y, \quad E_z = -\dfrac{\e_{xx} k_x + \e_{zx} k_z}{\eta}H_y.
    \label{eq:refraction_electricfield}
\end{equation}
Thus,
\begin{equation}
    r_h \equiv \dfrac{H_y^r}{H_y^i} = \dfrac{\e_{zz} k_z^i + \e_{xz} k_x}{\e_{zz} k_z^r + \e_{xz} k_x} \dfrac{E_x^r}{E_x^i} = \dfrac{\e_{zz} k_z^i + \e_{xz} k_x}{\e_{zz} k_z^r + \e_{xz} k_x} r_x.
\end{equation}
The relation between the transmission and the reflection coefficient for magnetic field is the same as for the tangnential component of the electric field,
\begin{equation}
    t_h = r_h +1.
\end{equation}

\subsection{Sample preparation and measurements}
\label{sec:experimental_verification}
\subsubsection{Numerical simulations}
For numerical simulations, the CST Microwave Studio 2019 software was used. Full-size numerical models of both samples of the metasurface were built. Numerically simulated isofrequency curves and H-field maps on top of both samples were calculated in eigenmode and transient solvers correspondently. 
\subsubsection{Experiment}
In the experiment, the excitation loop was fixed in the center of the metasurface sample and was connected to the first port of the two-port vector network analyzer (VNA). At the same time, a probe loop of the H field, with the same diameter as the excitation loop, was connected to the second port of the VNA. The manufactured sample was fixed using a foam substrate and was placed between the excitation loop and the probe for near-field measurements. The probe loop was fixed on the near-field 3D scanner across the plane parallel to the metasurface on the opposite side of the structure such that the gap between the probe and sample was equal to 7 mm.  After, measured H-field maps were converted to the isofrequency contours using space Fourier transformation.

The difference between numerical simulation and experimental results is due to the asymmetry of the excitation and probe loops, which results in the asymmetry of the field profiles, and sample imperfections.

\medskip

%
\bibliographystyle{MSP}

\end{document}